\documentclass{elsart}

\usepackage{natbib}
\usepackage{graphicx}
\usepackage{epstopdf}

\usepackage{amssymb}
\usepackage{pdflscape}

\begin{document}

\begin{frontmatter}



\title{The binary near-Earth asteroid (175706)~1996~FG$_3$ --- An observational constraint on its orbital evolution}


\author[a]{P.~Scheirich\corauthref{cor1}},
\corauth[cor1]{Corresponding author. Fax: +420 323 620263.\\}
\ead{petr.scheirich@centrum.cz}
\author[a]{P. Pravec},
\author[b,c,d]{S.A. Jacobson},
\author[e]{J. \v{D}urech},
\author[a]{P. Ku\v{s}nir\'{a}k},
\author[a]{K. Hornoch},
\author[f]{S. Mottola},
\author[f]{M. Mommert},
\author[f]{S. Hellmich},
\author[g]{D. Pray},
\author[h]{D. Polishook},
\author[i]{Yu.N. Krugly},
\author[j]{R.Ya. Inasaridze},
\author[j]{O.I. Kvaratskhelia},
\author[j]{V. Ayvazian},
\author[i]{I. Slyusarev},
\author[k]{J. Pittichov\'{a}},
\author[l]{E. Jehin},
\author[l]{J. Manfroid},
\author[l]{M. Gillon},
\author[m]{A. Gal\'{a}d},
\author[n]{J. Pollock},
\author[o,p]{J. Licandro},
\author[o,p]{V. Al\'{i}-Lagoa},
\author[q]{J. Brinsfield},
\author[r]{I.E. Molotov}.

\address[a]{Astronomical Institute, Academy of Sciences of the Czech Republic, Fri\v{c}ova 1, CZ-25165 Ond\v{r}ejov, Czech Republic}
\address[b]{Laboratoire Lagrange, Observatoire de la C\^{o}te d'Azur, Bd. De l'Observatoire, CS 34229, 06304 Nice Cedex 4, France}
\address[c]{Bayerisches Geoinstitut, Universt\"{a}t Bayreuth, 95440 Bayreuth, Germany}
\address[d]{Department of Astrophysical and Planetary Sciences, University of Colorado, Boulder, USA}
\address[e]{Astronomical Institute, Faculty of Mathematics and Physics, Charles University in Prague, V Hole\v{s}ovi\v{c}k\'ach 2, 18000 Prague, Czech Republic}
\address[f]{German Aerospace Center (DLR), Institute of Planetary Research, Rutherfordstr. 2, 12489, Berlin, Germany}
\address[g]{Sugarloaf Mountain Observatory, Massachusetts, USA}
\address[h]{Department of Earth, Atmospheric, and Planetary Sciences, Massachusetts Institute of Technology, Cambridge, MA 02139, USA}
\address[i]{Institute of Astronomy of Kharkiv National University, Sumska Str. 35, Kharkiv 61022, Ukraine}
\address[j]{Kharadze Abastumani Astrophysical Observatory, Ilia State University, K.Cholokoshvili Av. 3/5, Tbilisi 0162, Georgia}
\address[k]{Jet Propulsion Laboratory, Pasadena, CA 91109, USA}
\address[l]{Institut d'Astrophysique et de G\'{e}ophysique, Sart-Tilman, B-4000 Li\`{e}ge, Belgium}
\address[m]{Modra Observatory, Department of Astronomy, Physics of the Earth, and Meteorology, FMFI UK, Bratislava SK-84248, Slovakia}
\address[n]{Physics and Astronomy Department, Appalachian State University, Boone, NC 28608, USA}
\address[o]{Instituto de Astrof\'{i}sica de Canarias, c/v\'{i}a L\'{a}ctea s/n, 38200 La Laguna, Tenerife, Spain}
\address[p]{Departamento de Astrof\'{i}sica, Universidad de La Laguna, 38206, La Laguna, Tenerife, Spain}
\address[q]{Via Capote Observatory, Thousand Oaks, CA, USA}
\address[r]{Keldysh Institute of Applied Mathematics, RAS, Miusskaya sq. 4, Moscow 125047, Russia}


\newpage
Proposed running head: Binary near-Earth asteroid 1996 FG$_3$

\vspace{2cm}
Editorial correspondence to: \\
Peter Scheirich, Ph.D.        \\
Astronomical Institute AS CR \\
Fri\v{c}ova 1                \\
Ond\v{r}ejov                 \\
CZ-25165                     \\
Czech Republic               \\
Phone: 00420-323-620115      \\
Fax: 00420-323-620263        \\
E-mail address: petr.scheirich@centrum.cz \\

\newpage

\begin{abstract}

Using our photometric observations taken between April 1996 and January 2013 and other
published data, we derived properties of the binary near-Earth asteroid
(175706)~1996~FG$_3$ including new measurements constraining evolution of the mutual orbit with potential
consequences for the entire binary asteroid population. We also refined
previously determined values of parameters of both components, making 1996~FG$_3$
one of the most well understood binary asteroid systems.
With our 17-year long dataset, we determined the orbital vector with a substantially
greater accuracy than before and we also placed constraints on a stability of the orbit.  Specifically,
the ecliptic longitude and latitude of the orbital pole are $266^{\circ}$ and $-83^{\circ}$, respectively, with
the mean radius of the uncertainty area of $4^{\circ}$, and the orbital period is $16.1508 \pm 0.0002$~h
(all quoted uncertainties correspond to 3$\sigma$).
We looked for a quadratic drift of the mean anomaly of the satellite and obtained a value of $0.04 \pm 0.20$~deg/$\mbox{yr}^2$, i.e.,
consistent with zero.
The drift is substantially lower than predicted by the pure binary YORP (BYORP) theory of McMahon and Scheeres (McMahon, J., Scheeres, D. [2010]. Icarus 209, 494--509)
and it is consistent with the theory of an equilibrium between BYORP and tidal torques for synchronous binary asteroids as proposed by Jacobson and Scheeres (Jacobson, S.A., Scheeres, D. [2011]. ApJ Letters, 736, L19).
Based on the assumption of equilibrium, we derived a ratio of the quality factor and tidal Love number of $Q/k = 2.4 \times 10^5$ uncertain by a factor of five.
We also derived a product of the rigidity and quality factor of $\mu Q = 1.3 \times 10^7$~Pa using the theory that assumes an elastic response of the
asteroid material to the tidal forces.  This very low value indicates that the primary of 1996~FG$_3$ is a `rubble pile', and it also calls for
a re-thinking of the tidal energy dissipation in close asteroid binary systems.
\end{abstract}

\begin{keyword}
Asteroids, dynamics; Near-Earth objects; Photometry


\end{keyword}

\end{frontmatter}

\newpage

\section{Introduction}
\label{Introduction}

The near-Earth asteroid (175706) 1996~FG$_3$ (hereafter referred to as FG3) was discovered by R.H.~McNaught from Siding Spring on 1996 March 24.
Its binary nature was revealed by Pravec et al.~(1998).  The asteroid was observed thoroughly with
a variety of techniques during six apparitions from 1996 to 2013.  Our photometric dataset represents
the longest coverage obtained for a binary near-Earth asteroid so far, providing a unique opportunity
to study an evolution of the mutual orbit of a small binary asteroid.
Moreover, a detailed understanding of this asteroid system is motivated by its accessibility from the Earth
with a delta-V requirement of only 5.16 km/s (Binzel et al., 2004)  and so it is a popular candidate for past and future proposed space missions.


There have been a number of recent observations
and models of this binary published in the literature, however rather than reviewing
them here, we will delay their review until after we present our new data. Then in
Section~\ref{ParamsSect}, we can present and discuss all of the system parameters including
those previously published and those updated from our work.

The structure of this paper is as follows.  In Section~\ref{Model}, we present a model of the mutual orbit of the components of
FG3 constructed from our complete photometric dataset 1996--2013.  Then in Section~\ref{ParamsSect}, we summarize all known parameters
of the binary, updating them using our new data where needed.  In Section~\ref{byorp}, we then discuss implications of the current known
characteristics of the binary, especially on the BYORP theory from the derived low upper limit on the binary's orbital drift.

\section{Model of the mutual orbit} \label{Model}
\subsection{Observational data} \label{ObsData}

\renewcommand{\arraystretch}{1.3}
\begin{table} \caption{Observations}\label{TableData}
\begin{center}
  \begin{tabular}{cccc}
\hline
Time span                 &  No. of nights  & Telescope    & References \\
\hline
1996-04-09 to 1996-04-21  &  10             &  0.6-m Bochum         &  ML00 \\
1998-12-03 to 1999-01-09  &  8              &  0.65-m Ond\v{r}ejov   &  P00 \\
                          &  4              &  1.23-m Calar Alto     &  ML00 \\
                          &  3              &  0.7-m Kharkiv        &  P00 \\
                          &  3              &  0.61-m TMO &  P00 \\
                          &  1              &  1.55-m Catalina       &  P00 \\
2009-04-12 to 2009-04-17  &  4              &  0.65-m Ond\v{r}ejov   & This work \\  
                          &  1              &  0.5-m Carbuncle Hill & This work \\   
2010-12-14 to 2011-02-09  &  5              &  1.23-m Calar Alto     & This work \\
                          &  3              &  3.5-m Apache Point    & This work \\
                          &  2              &  2.2-m UH & This work \\
2011-11-23 to 2012-02-12  &  4              &  0.5-m Sugarloaf Mt.  & This work \\
                          &  3              &  0.46-m Wise Obs. & This work \\ 
                          &  3              &  0.7-m Abastumani  & This work \\
                          &  3              &  2.1-m Kitt Peak   & This work \\
                          &  2              & 0.7-m Kharkiv                 & This work \\ 
                          &  2              & 0.65-m Ond\v{r}ejov           & This work \\ 
                          &  2              & 0.6-m TRAPPIST                & This work \\
                          &  1              & 0.6-m Modra                   & This work \\ 
                          &  1              & 0.41-m PROMPT                 & This work \\ 
                          &  1              & 0.36-m Via Capote             & This work \\ 
                          &  1              & 0.82-m IAC-80                 & This work \\ 
2013-01-05 to 2013-01-07  &  3              & 2.1-m Kitt Peak               & This work \\
\hline
 \end{tabular}
\end{center}
References:
ML00 (Mottola and Lahulla, 2000),
P00 (Pravec et al., 2000),
\end{table}

\renewcommand{\arraystretch}{1.3}
\begin{table} \caption{Observational stations}\label{TableObs}
\begin{center}
  \begin{tabular}{llc}
\hline
Telescope            &  Observatory                                    & References for \\
                     &                                                 & observational and \\
                     &                                                 & reduction procedures \\
\hline
3.5-m Apache Point   & Apache Point Observatory, New Mexico, USA       &  1 \\
2.2-m UH             & University of Hawaii, USA                       &  2 \\
2.1-m Kitt Peak      & Kitt Peak, Arizona, USA                         &  1 \\
1.55-m Catalina      & Catalina Observatory, Arizona, USA              &  P00 \\
1.23-m Calar Alto    & Calar Alto, Spain                               &  ML00 \\
0.82-m IAC-80        & Instituto de Astrof\'{i}sica de Canarias, Spain & TR10 \\
0.7-m Abastumani     & Abastumani Astrophysical Observatory, Georgia   & K02, Pi12 \\
0.7-m Kharkiv        & Kharkiv National University, Ukraine            &  1998/1999: K02, \\
                     &                                                 &  2011/2012: P12\\
0.65-m Ond\v{r}ejov  & Ond\v{r}ejov, Czech Republic                    &  1998/1999: P00,\\
                     &                                                 &  2009-2012: P06 \\
0.61-m TMO           & Table Mountain Observatory, California, USA     &  P00 \\
0.60-m Bochum        & La Silla, Chile                                 &  ML00 \\
0.6-m Modra          & Modra, Slovakia                                 & G07 \\
0.6-m TRAPPIST       & La Silla, Chile                                 & P14 \\
0.5-m Carbuncle Hill & Carbuncle Hill Observatory, Massachusetts, USA  &  WP09 \\
0.5-m Sugarloaf Mt.  & Sugarloaf Mountain Observatory, Massachusetts   &  W13 \\
0.46-m Wise Obs.     & Wise Observatory, Israel                        &  B08, PB09 \\
0.41-m PROMPT        & Cerro Tololo, Chile                             & P12 \\
0.36-m Via Capote    & Via Capote Observatory, California, USA         & P12 \\
\hline
 \end{tabular}
\end{center}
References:
1 -- The data were reduced using IRAF and the PHOT package, following the methods by Harris and Lupishko (1989).
2 -- The observations were obtained using Tektronix $2048\times 2048$ CCD camera at the f/10 focus of the telescope (image scale of 0.219" pixel$^{-1}$)
through Kron-Cousins filter R.\\
B08 (Brosch et al., 2008),
G07 (Gal\'ad el al., 2007),
K02 (Krugly et al., 2002),
ML00 (Mottola and Lahulla, 2000),
P00 (Pravec et al., 2000),
P12 (Pravec et al., 2012, Supplementary Material),
P14 (Pravec et al. 2014),
PB09 (Polishook and Brosch, 2009),
Pi12 (Pilcher et al. 2012),
TR10 (Trigo-Rodr\'{\i}guez et al., 2010),
W13 (Warner et al., 2013),
WP09 (Warner and Pray, 2009).
\end{table}

The data used in our analysis, obtained during six apparitions, are summarized in Table~\ref{TableData}.
The references and descriptions of observational procedures of the individual observatories are summarized in Table~\ref{TableObs}.

The data were reduced using the standard technique described in Pravec et al. (2006).
By fitting a two-period Fourier series to data points outside mutual (occultation or eclipse) events, the rotational lightcurves of the
primary (short-period) and the secondary (long-period), which are additive in intensities, were separated.
The long-period component containing the mutual events and the secondary rotation lightcurve is
then fitted in subsequent numerical modeling (Sect.~\ref{NumModel}).

A special approach was needed for the 2013 data where the primary rotational lightcurve was incompletely described by the actual observations.
Using a shape and rotation model for the primary that we obtained from data from the previous apparitions, we generated a synthetic
rotational lightcurve of the primary and subtracted it from the data to obtain the long-period component of the lightcurve on the three nights
in January 2013.


\subsection{Numerical model} \label{NumModel}

We have constructed a model of the binary using the technique of Scheirich and Pravec (2009) that we further developed and enhanced in this work.
In the following, we outline the basic points of the method, but we refer the reader to the 2009 paper for details of the technique.  We also describe
new features and improvements of the method that we developed recently.

The shapes of the components were represented with ellipsoids, orbiting each other on
a Keplerian orbit with apsidal precession, and allowing for a quadratic drift in mean anomaly.

The primary was modeled as an oblate spheroid, with its spin axis assumed to be normal to the mutual orbital plane of the components.
The shape of the secondary was taken as a prolate spheroid in synchronous rotation, with its
long axis aligned with the centers of the two bodies.  The shapes were approximated with 1016 and 252 triangular facets for the primary
and the secondary, respectively.  The components were assumed to have the same albedo.
The brightness of the system as seen by the observer was computed as a sum of contributions from all visible facets using a
ray-tracing code that checks which facets are occulted by or are in shadow from the other body.
A combination of Lommel-Seeliger and Lambert scattering
laws was used (see, e.g., Kaasalainen et al., 2002).

The quadratic drift in mean anomaly, $\Delta M_d$, was fitted as an independent parameter.
It is the coefficient in the second term of the expansion of the time-variable mean anomaly:
\begin{equation}
M (t) = M (t_0) + n (t-t_0) + \Delta M_d (t - t_0)^2, \label{dMd1}
\end{equation}
where
\begin{equation}
\Delta M_d = \frac{1}{2} \dot{n}, \label{dMd2}
\end{equation}
where $n$ is the mean motion, $t$ is the time, and $t_0$ is the epoch.

$\Delta M_d$ was stepped from $-3$ to $+3$~deg/yr$^2$ with a step of 0.02~deg/yr$^2$, and all other parameters were fitted at each step.

We also tested a possibility of a larger value of $\Delta M_d$ that would cause an integer change of the number of orbital cycles between the 1998-1999 and the
2009 apparitions, but we found such large orbital drifts entirely incompatible with the data.

To reduce the complexity of the model, we estimated an upper limit on the eccentricity of the mutual orbit by fitting
the data from the best covered apparition 2011-2012.
The model includes a precession of the line of apsides.
The pericenter drift rate depends on the polar flattening of the primary (see Murray and Dermott, 1999, Eq. (6.249)), but the
polar flattening is poorly constrained from the data (see Table~\ref{tableProp}),
so instead we fit the drift rate as an independent parameter.
Its initial values were stepped in a range from zero to $35^{\circ}/{\rm day}$. This range encompasses all
plausible values for the flattening of the primary and other parameters of the system.

Since the upper limit on eccentricity of the mutual orbit was found to be low ($e_{\rm max}=0.07$), in modeling data from all apparitions
together, we set the eccentricity equal to zero for simplicity.  This assumption had a negligible effect on the accuracy of other derived
parameters of the binary model.\footnote{We tested an effect of the zero eccentricity assumption with a following experiment:
We generated synthetic lightcurve data for the binary system while forcing its eccentricity to be 0.07.  Then we modeled the data with the assumption
of circular orbit.  We found that there were only minor or negligible differences in the fitted parameters; the largest difference of 5\% was
in the semimajor axis.}

\begin{figure}[h]
\begin{center}
\includegraphics[width=\textwidth]{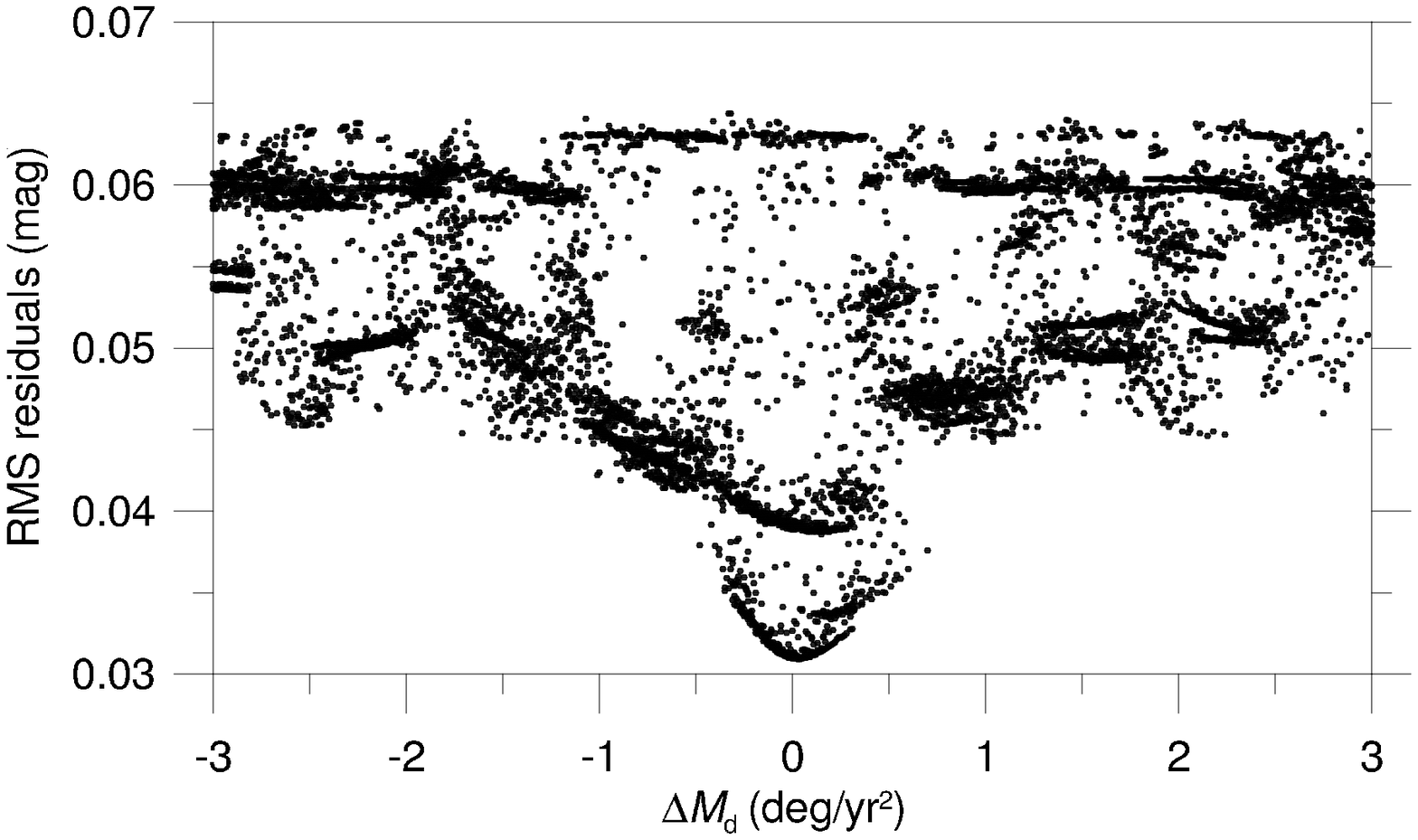}
\end{center}
\caption{\small The RMS residuals vs. $\Delta M_d$ for solutions of the model presented in Section~\ref{NumModel}. Each dot represents the best-fit result with $\Delta M_d$ fixed and other parameters varied.}
\label{RMS_vs_DMd}
\end{figure}

Across all observations, we found a unique solution for the system parameters, see
Table 3. We describe and discuss these parameters in Section~\ref{ParamsSect}.
A plot of RMS residuals vs. $\Delta M_d$ is shown in Fig.~\ref{RMS_vs_DMd}.

\begin{figure}[h]
\begin{center}
\includegraphics[width=\textwidth]{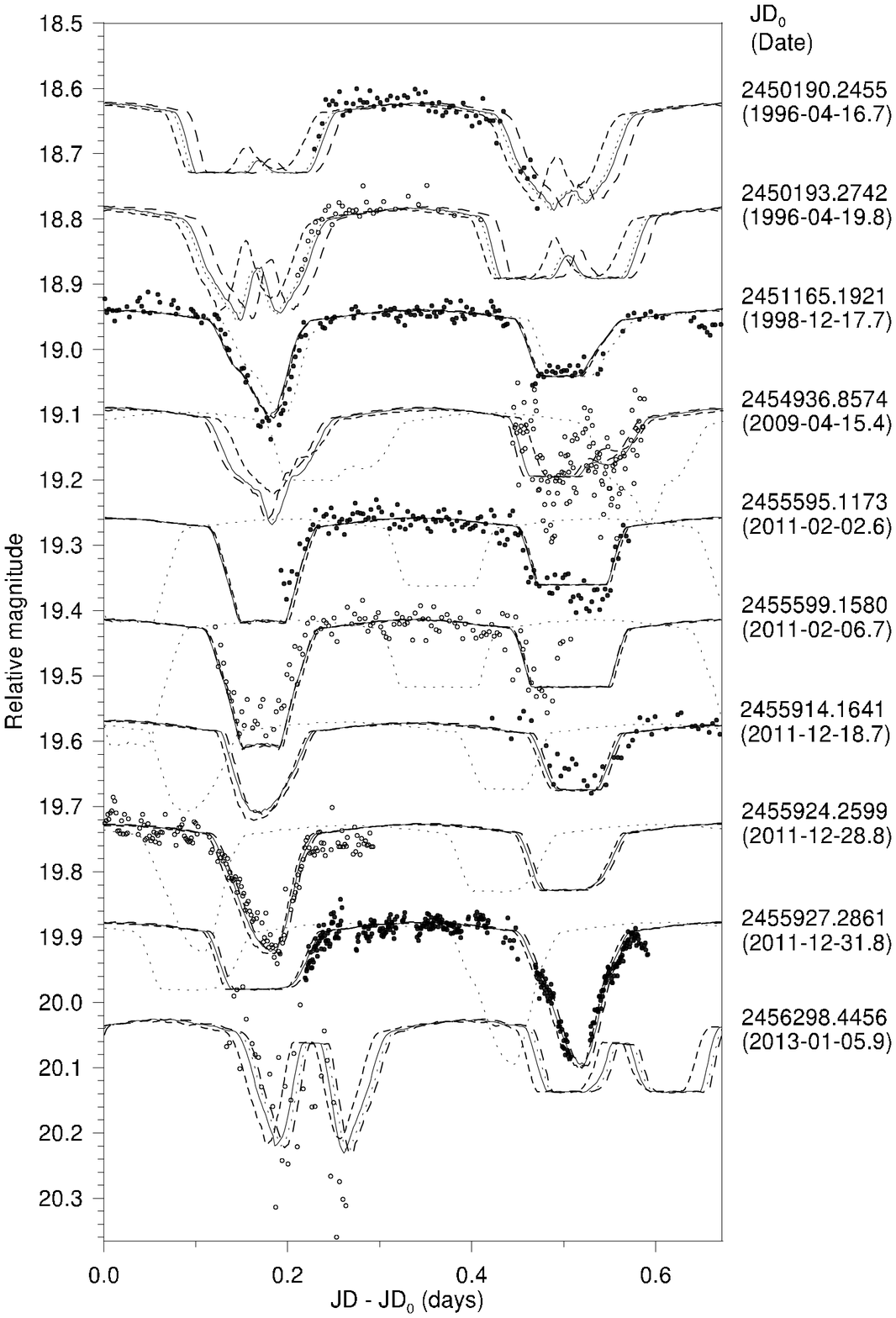}
\end{center}
\caption{\small Selected data of the long-period lightcurve component of 1996 FG$_3$. The observed data are marked as points. The solid curve represents the synthetic lightcurve of the best-fit solution.
The dashed curves are solutions at the 3-$\sigma$ uncertainty in $\Delta M_d$ (short dashes: $\Delta M_d$ equal to $+0.24$, long dashes: $\Delta M_d$ equal to $-0.16$~deg/yr$^2$).
For comparison, the dotted curve represents a model with $\Delta M_d = -1.30$~deg/yr$^2$ as predicted by the pure BYORP theory without tides (see Section~\ref{byorp}).}
\label{1996fg3_96-98-a9-11-12_synth}
\end{figure}

\begin{figure}[h]
\begin{center}
\includegraphics[width=\textwidth]{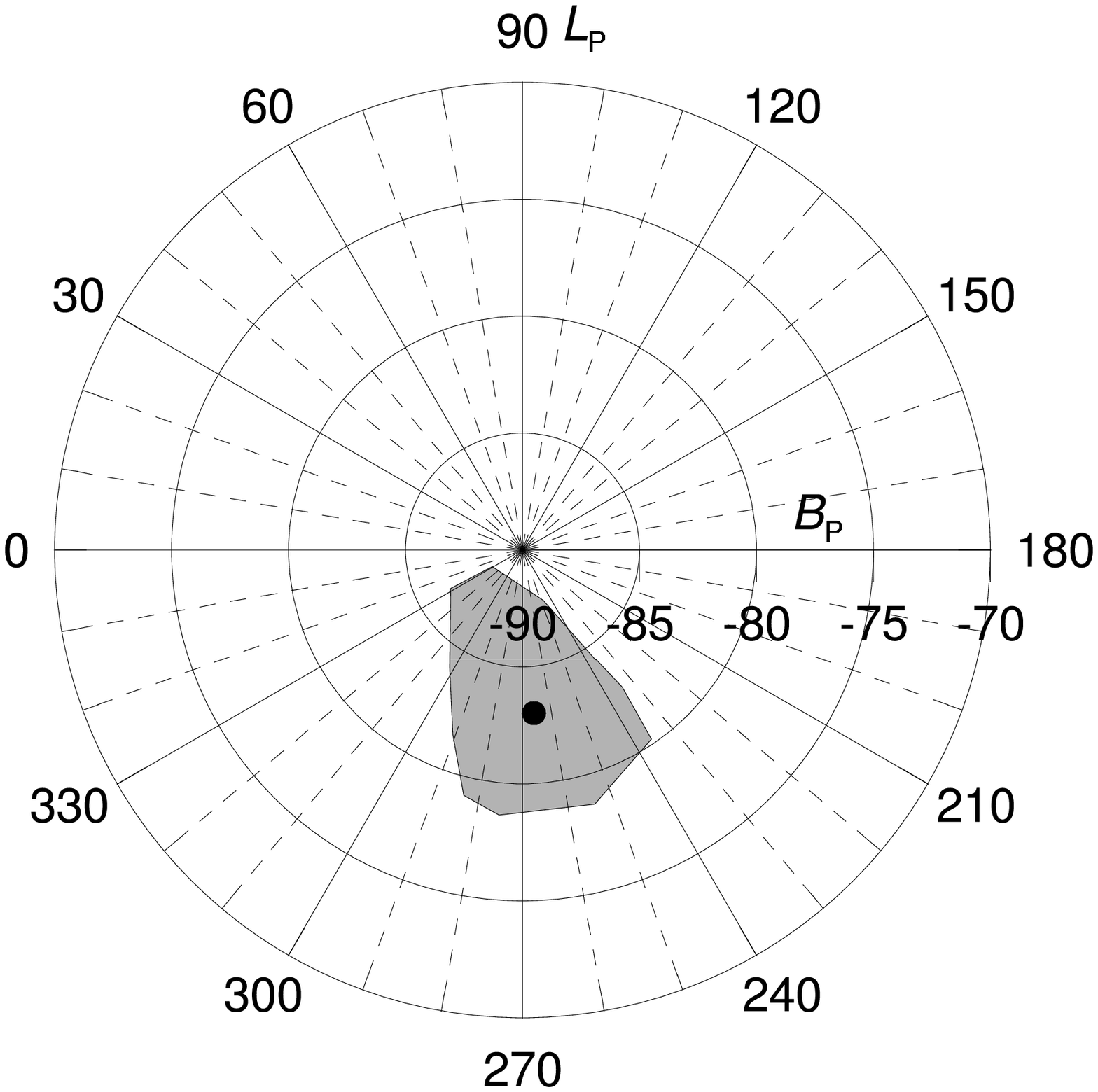}
\end{center}
\caption{\small Area of admissible poles for the mutual orbit in ecliptic coordinates (grey area). The dot is the nominal solution given in Table~\ref{tableProp}.
This area corresponds to $3\sigma$ confidence level.}
\label{1996fg3_LB_polar}
\end{figure}

Examples of the long-period component data together with the synthetic
lightcurve of the best-fit solution are presented in Fig.~\ref{1996fg3_96-98-a9-11-12_synth}.
An uncertainty area of the orbital pole is shown in Fig.~\ref{1996fg3_LB_polar}.

We estimated realistic uncertainties of the fitted parameters using the procedure described in Scheirich and Pravec (2009).
For each parameter, we obtained its admissible range that corresponds to a 3-$\sigma$ uncertainty.
Since the 2013 data required a special approach (see Section~\ref{ObsData}), we examined the sensitivity of the results on them and found that the values obtained and their
errors are dominated by the previous apparitions. The 2013 data, however, show a good agreement with the results obtained from the previous epochs and they increase confidence in the results.
As shown in Fig.~\ref{1996fg3_96-98-a9-11-12_synth}, an apparent discrepancy of the 2009 data is also worth noting.
There is an apparent double secondary event in the data, which is not fully reproduced with the nominal solution.
This event occurred, however, in an almost tangent geometry in which its shape is extremely sensitive to the actual figure of the primary,
which the model assumes is an oblate spheroid.
The data of 2009 are also of much lower quality than the data from other apparitions, so the splitting of the event may possibly be an observational artifact.

We also attempted to model the shape and spin of the primary using the lightcurve inversion method of Kaasalainen and Torppa (2001) and Kaasalainen et al. (2001)
with the data for the primary lightcurve (outside the mutual events and with the secondary's rotational component subtracted)
from the five apparitions 1996--2012.  We found that a unique solution for the shape and pole of the primary could not be obtained,
with a wide range of possible shapes and poles that produced equally good fits to the data.
Apparently, the ambiguity is due to the combination of two features: (1) The primary's shape irregularities are low, the shape does not differ from oblate spheroid much.
(2) Due to the asteroid's obliquity being close to 180 degrees, the object was observed at near-equator on aspects only.  Assuming zero inclination
of the mutual orbit of the components to the primary's equator, i.e., the primary's pole being the same as the orbital pole, we obtained
a unique solution for the primary rotational period of $3.595195 \pm 0.000003$~h (3-$\sigma$ uncertainty; this includes also the uncertainty
of the orbital pole of 4 degrees).  Nevertheless, even for the constrained pole, the primary's shape was not derived uniquely,
as the polar flattening of the primary is poorly defined with the observations at near-equator on aspects.

\newpage


\section{Parameters of 1996~FG$_3$}
\label{ParamsSect}

In this section, we summarize known parameters of the binary asteroid.  We overview previous publications and we also derive or refine some
parameters using our newest data.  In Table~\ref{tableProp}, we provide the best estimated values for the parameters of FG3.

\renewcommand{\arraystretch}{1.3}
\begin{table} \caption{Properties of binary asteroid (175706) 1996 FG$_3$.}\label{tableProp}
\begin{center}
  \begin{tabular}{cccc}
\hline
Parameter                                  &  Value                       &  Unc.    &  Reference\\
\hline
\multicolumn{4}{l}{Whole system:} \\
$H_V$                                      & $17.833 \pm 0.024$           & 1$\sigma$ & W11 \\
$G$                                        & $-0.041 \pm 0.005$           & 1$\sigma$ & W11 \\
$V - R$                                    & $0.380 \pm 0.003$            & 1$\sigma$ & P00 \\
$R - I$                                    & $0.334 \pm 0.003$            & 1$\sigma$ & P00 \\
$B - V$                                    & $0.708 \pm 0.005$            & 1$\sigma$ & P00 \\
$D_{\rm eff}$ (km)                         & $1.71 \pm 0.07$              & 1$\sigma$ & W11 \\
$p_V$                                      & $0.044 \pm 0.004$            & 1$\sigma$ & W11 \\
$\Gamma$ (J~m$^{-2}$~K$^{-1}$~s$^{-1/2}$)  & $120 \pm 50$                 & 1$\sigma$ & W11 \\
Taxon. class                               & C, B, Xc                     &           & B01, W12, P13 \\
Spectrum                                   & \multicolumn{2}{l}{Featureless and flat in visible range.}                 & B01 \\
                                           & \multicolumn{2}{l}{Shallow features near 1.2 and 2.0 $\mu$m.}             & W12 \\
                                           & \multicolumn{2}{l}{Abs. band of OH or waterbearing mineral.}      & R13 \\
Meteorite analogue                         & CM2, CM, C2                  &           & L11, P12, P13 \\
\hline
\multicolumn{4}{l}{Primary:}  \\
$D_{\rm 1,C}$ (km)                         & $1.64 \pm 0.20$              & 3$\sigma$ & this work \\
$D_{\rm 1,V}$ (km)                         & $1.69^{+0.24}_{-0.21}$       & 3$\sigma$ & this work\\
$P_1$ (h)                                  & $3.595195 \pm 0.000003$      & 3$\sigma$ & this work\\
$(A_1 B_1)^{1/2}/C_1$                      & $1.2^{+0.5}_{-0.2}$          & 3$\sigma$ & this work\\
$A_{\rm 1}/B_{\rm 1}$                      & $1.06 \pm 0.03$              & 3$\sigma$ & P06 \\
$\rho_1 = \rho_2$ (g~cm$^{-3}$)            & $1.3 \pm 0.5$                & 3$\sigma$ & this work\\
\hline
\multicolumn{4}{l}{Secondary:}  \\
$D_{\rm 2,C}/D_{\rm 1,C}$                  & $0.29 \pm 0.02$              & 3$\sigma$ & this work\\
$D_{\rm 2,C}$ (km)                         & $0.48 \pm 0.07$              & 3$\sigma$ & this work \\
$D_{\rm 2,V}$ (km)                         & $(0.49 \pm 0.08)$            & 3$\sigma$ & this work \\
$P_2$ (h)                                  & $16.15 \pm 0.01$             & 1$\sigma$ & P06\\
$A_{\rm 2}/B_{\rm 2}$                      & $1.3 \pm 0.2$                & 3$\sigma$ & this work\\
\hline
\multicolumn{4}{l}{Mutual orbit:} \\
$a/D_{\rm 1,C}$                            & $1.5^{+0.3}_{-0.2}$          & 3$\sigma$ & this work\\
$(L_{\rm P}, B_{\rm P})$ (deg.)            & $(266, -83) \pm 4^a$           & 3$\sigma$ & this work\\
$P_{\rm orb}$ (h)                          & $16.1508 \pm 0.0002$         & 3$\sigma$ & this work\\
$e_{\rm max}$                              & 0.07                         & 3$\sigma$ & this work\\
$\Delta M_d$ (deg/yr$^2$)                  & $0.04 \pm 0.20$              & 3$\sigma$ & this work\\
$\dot{P}_{\rm orb}$ (h/yr)                 & $-0.000007 \pm 0.000033 $    & 3$\sigma$ & this work\\
$\dot{a}$ (cm/yr)                          & $-0.07 \pm 0.34$             & 3$\sigma$ & this work\\
\hline
 \end{tabular}
\end{center}
References: B01 (Binzel et al., 2001),
L11 (de Le\'on et al., 2011),
P00 (Pravec et al., 2000),
P06 (Pravec et al., 2006),
P12 (Popescu et al., 2012),
P13 (Perna et al., 2013),
R13 (Rivkin et al., 2013),
W11 (Wolters et al., 2011),
W12 (Walsh et al., 2012).\\
$^a$ This is the mean radius of the uncertainty area; see its actual shape in Fig.~\ref{1996fg3_LB_polar}.
\end{table}

In the first part of the table, we present data derived from optical, thermal and spectroscopic observations of the system.
$H_V$ and $G$ are the mean absolute magnitude and the phase parameter of the $H$--$G$ phase relation (Bowell et al., 1989),
$V-R$, $R-I$ and $B-V$ are the asteroid's color indices in the Johnson-Cousins photometric system,
$D_{\rm eff}$ is the effective diameter of system,
$p_{\rm V}$ is the visual geometric albedo and $\Gamma$ is the thermal inertia of the asteroid surface.

The best values for these parameters, except for the color indices that were measured by Pravec et al. (2000), were obtained by Wolters et al. (2011).
They took thermal and visual observations of the asteroid using the ESO VLT and NTT, respectively.  They combined their visual photometric data
with measurements by Pravec et al. (2000) and Mottola and Lahulla (2000) and, thanks also to their measurement taken at the small solar phase of 1.4 deg,
they obtained the most precise value for the mean absolute magnitude of the system.  The previous $H_V$ value by Pravec et al. (2000) is in agreement
to about $2 \sigma$.  We have checked it also with our recent absolute photometry from Ond\v{r}ejov in April 2009 and December 2011 (Table~1),
and we found an agreement with the $H, G$ values by Wolters et al. (2011) to a few hundredths of magnitude.  With the precise $H, G$ values, Wolters et al. (2011)
then applied the Advanced Thermophysical Model (ATPM) to their thermal observations and derived the effective diameter, geometric albedo and thermal inertia.
They assessed realistic uncertainties of the derived parameters.

The effective diameter and geometric albedo were derived also in two other recent works Mueller et al. (2011) and Walsh et al. (2012).
Mueller et al. reported $p_V = 0.042^{+0.035}_{-0.017}$ and $D_{\rm eff} = 1.84^{+0.56}_{-0.47}$
and Walsh et al. derived $p_V = 0.039 \pm 0.012$ and $D_{\rm eff} = 1.90 \pm 0.28$~km.  Their values are in agreement with those by Wolters et al. (2011),
but they are less precise, because both teams observed the asteroid at high phase angles $> 50^{\circ}$ and they used the NEATM in their modeling.
The NEATM is less accurate than the ATPM, and for data obtained at high phase angles it systematically overestimates diameter and underestimates albedo (Wolters and Green, 2009).
We adopt the values derived by Wolters et al. (2011) as the most precise data for these parameters.


In the most recent work on this topic, Yu et al. (2014) combined the observational data from the already mentioned publications and other sources.
From this dataset, they attempted to further refine these photometric and thermal parameters.
While the thermal modeling part of their paper is well done, their work suffers because they did not
properly assess the real uncertainties of their spin and shape model, which they constructed from an extremely limited dataset.
Though their derived values for the photometric and thermal parameters are close to those of Wolters et al. (2011),
we hesitate to utilize their values due to the incomplete assessment of their uncertainties.
Thus, we continue to use the parameters of Wolters et al. (2011), where the uncertainties were properly estimated.

Several works have been published reporting spectroscopic observations of FG3 in the visible and near-infrared spectral range, see the references in Table~\ref{tableProp}.
The obtained data generally agree with featureless and flat spectrum in the visible range. There are differences in the near-infrared range and some authors report
shallow features indicating presence of olivine, pyroxene and OH or water-bearing minerals.

Although there is no consensus on taxonomic classification of FG3 in the literature (Binzel et al., 2001; de Le\'on et al., 2011; Walsh et al., 2012),
they agree that the asteroid is composed of primitive material, and all the proposed taxonomic classes are consistent with the measured low geometric albedo.

In the next two parts of Table~\ref{tableProp}, we give parameters for the components of the binary.
The indices 1 and 2 refer to the primary and the secondary, respectively.
$D_{i,{\rm C}}$ is the cross-section equivalent diameter, i.e., the diameter of a sphere with the same cross section, of the $i$-th component
at the observed, i.e., equator-on aspect.
$D_{i,{\rm V}}$ is the volume equivalent diameter, i.e., the diameter of a sphere with the same volume, of the $i$-th component.
$D_{2,{\rm C}}/D_{1,{\rm C}}$ is the ratio between the cross-section equivalent diameters of the components.
$P_i$ is the rotational period of the $i$-th component.
$(A_1 B_1)^{1/2}/C_1$ is a ratio between the mean equatorial and the polar axes of the primary.
$A_i/B_i$ is a ratio between the equatorial axes of the $i$-th component (equatorial elongation).
$\rho_1 = \rho_2$ are the bulk densities of the two components, which we assumed to be the same in our modeling.

Most of the quantities were parameters of our model given in Section~\ref{NumModel} and we derived them from our observations.  The cross-section and
volume equivalent diameters of the components were derived using the $D_{\rm eff}$ value from Wolters et al. (2011) for the absolute size calibration
of our model, assuming the same geometric albedo for both components.  The uncertainties for all the parameters are realistic, corresponding to $3 \sigma$,
except for $D_{2,\rm V}$ where there may be present a systematic error due to the assumed shape of prolate ellipsoid ($B_2 = C_2$), so we give the value
in parentheses.

\begin{figure}[h]
\begin{center}
\includegraphics[width=\textwidth]{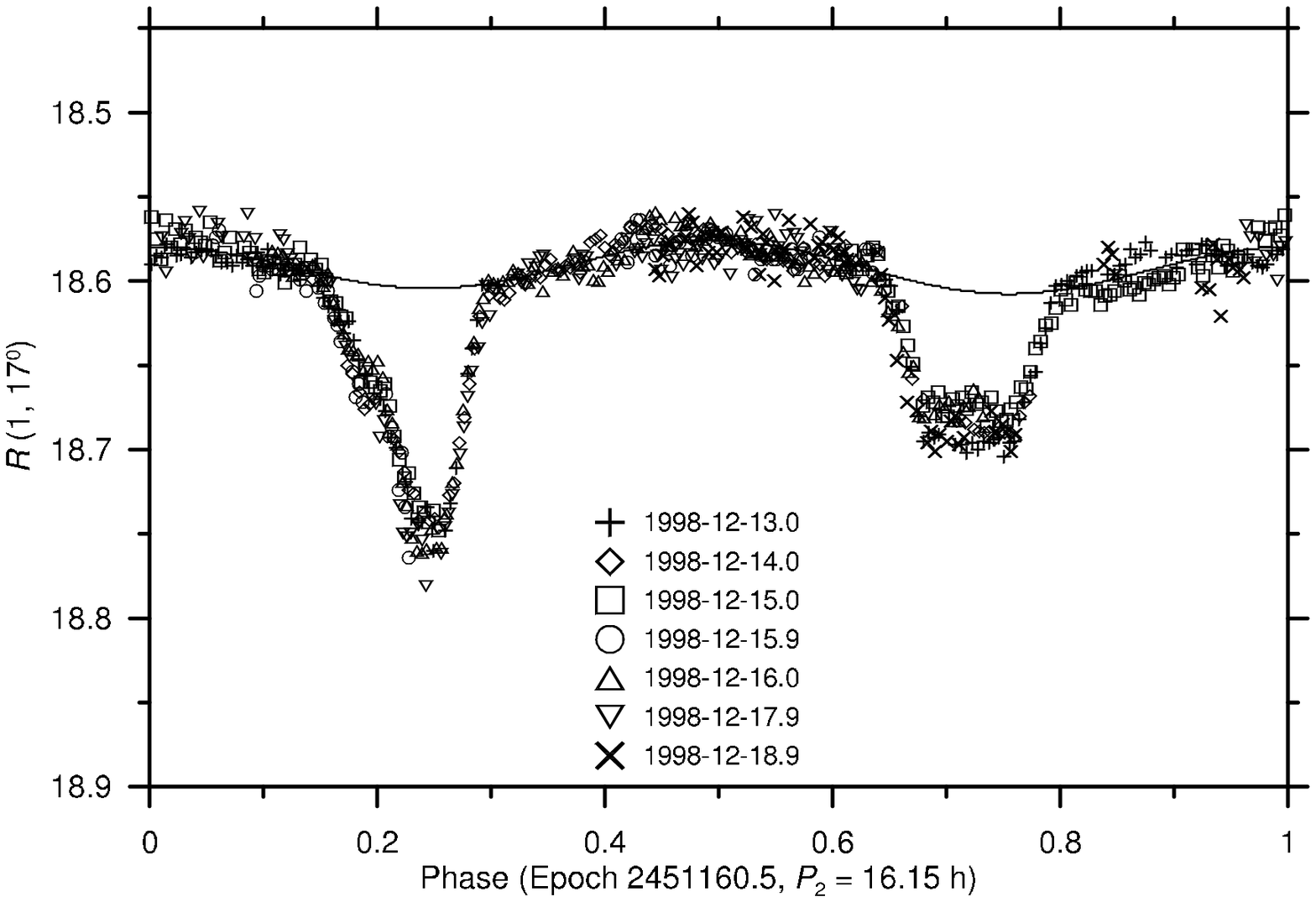}
\end{center}
\caption{\small The secondary's rotational lightcurve is best seen in the highest quality data taken by Mottola and Lahulla~(2000) and Pravec et al.~(2000) during
1998 December 13--18.  The curve is the 2nd order Fourier series fitted to the data outside the mutual events, plotted here to facilitate visibility of
the secondary lightcurve minima.  The primary's rotation was subtracted from the data before plotting.}
\label{1996fg3_1998_p2}
\end{figure}

As the rotational state of the secondary is particularly important for the interpretations we present in Section~\ref{byorp}, we have carefully re-analyzed data for
the secondary's rotational lightcurve.  The highest quality data on the secondary's rotation were taken during six nights 1998 December 13.0--18.9, with
four runs by Mottola and Lahulla (2000) and three runs by Pravec et al. (2000).  We present them in Fig.~\ref{1996fg3_1998_p2}.  They show that the long
axis of the secondary was approximately aligned with the centers of the two bodies; the lightcurve minima occurred during the eclipse/occultation
mutual events.  (The apparent small time offset, with the minima occurring about 0.4~h after the mutual event center, may be due to a libration of the secondary
or it may be a secondary shape effect.)  Other data with sufficient length and photometric quality taken in this and other apparitions 
were also consistent with the secondary's lightcurve minima occurring always near the times of mutual events; we have never observed a secondary rotational
lightcurve minimum occurring more than a couple ten degrees in mean anomaly from the mutual events.  This indicates that the secondary is in synchronous
rotation.  A moderate libration is possible, but the available data are not sufficient to derive it.

The asteroid FG3 was observed with the Arecibo and Goldstone radars by Benner et al. (2012).  They reported a rounded, slightly elongated shape of the primary
with diameter of about 1.8~km, with features along its equator resembling that of the equatorial ridge of binary asteroid (66391) 1999 KW$_4$
(Ostro et al., 2006).  They also mentioned that their data suggested a synchronous rotation of the secondary.  These preliminary estimates from the
radar observations are in agreement with our data, especially considering that their estimated primary diameter corresponds to the equatorial rather
than volume-equivalent diameter.

In the last part of Table~\ref{tableProp}, we summarize the parameters of the mutual orbit of the binary components.
$a$ is the semimajor axis, $L_{\rm P}, B_{\rm P}$ are the ecliptic coordinates of the orbital pole in the equinox J2000, $P_{\rm orb}$ is the orbital period,
$e_{\rm max}$ is the upper limit on eccentricity, and $\Delta M_d$ is the quadratic drift in mean anomaly.
We also give the time derivatives of the orbital period and the semimajor axis, derived from $\Delta M_d$.

Earlier works where some of the binary parameters were derived are Pravec et al. (2000), Mottola and Lahulla (2000), Pravec et al. (2006) and
Scheirich and Pravec (2009).  Their results are generally in agreement with our current best estimated parameters, though naturally their values were
less precise because they were derived from the observations from the first two apparitions only.

The binary (175706) 1996~FG$_3$ appears to be a typical near-Earth binary asteroid according to most of its parameters.
It is an apparent outlier in only one thing,
plus related quantities: its primitive taxonomic type and composition.
Most known NEA binaries are S and other rocky types.
However, it may be only a bias due to the well known observational selection effect against dark-albedo NEAs; primitive-type binaries might be
actually as common as those of S and similar types.
Probably related to its primitive composition is also its relatively low bulk density of $\sim 1.3$~g~cm$^{-3}$.
A consequence of it appears to be also its relatively long primary rotation period of 3.6~h; the primary periods of NEA binaries concentrate
between 2.2 and 2.8 h (Pravec et al., 2006), but the FG3 primary rotates more slowly for the lower critical spin frequency for its low bulk density.
Indeed, the normalized total angular momentum content of FG3 is $\alpha_L = 1.00 \pm 0.08$ (1-$\sigma$ uncertainty),
i.e., in the range 0.9--1.3 for small near-Earth and main belt asteroid binaries and exactly as expected for the proposed formation of small binary asteroids by
fission of critically spinning rubble-pile progenitors (Pravec and Harris, 2007).

\newpage

\newpage

\newpage

\newpage
\section{Implications for the BYORP effect and tidal parameters}\label{byorp}

While most of the parameters of the binary FG3 were known, albeit with lower accuracy, from earlier publications, we study the secular change of
the mutual orbit for the first time.
We have found that there is a very low or zero drift of the orbit over the
observational interval of 17 years.  It is interesting to compare this observation with theory of the binary YORP (BYORP) effect.


The BYORP effect is a secular change of the mutual orbit of a binary asteroid system with a synchronous satellite due to the emission of thermal radiation
from the asymmetric shape of that satellite.
It was first hypothesized by \'{C}uk and Burns (2005) after the prediction of two other radiative torques: the Yarkovsky and YORP
effects (see the review by Bottke et al., 2006).
Both of these torques have been observed acting on asteroid systems as expected according to the theory: the Yarkovsky effect
was first detected by Chesley et al. (2003) on 6489 Golevka and the YORP effect was
first detected by Lowry et al. (2007) and Taylor et al. (2007) on 54509 (2000 PH$_5$),
and Kaasalainen et al. (2007) on (1862) Apollo.
The BYORP effect is identical to the YORP effect
except the lever arm of the radiative torque extends from the surface elements of the
satellite to the center of mass of the mutual orbit rather than to the center of mass of
the asteroid itself. Using the YORP effect as a guide, McMahon and Scheeres
(2010a,b) built a detailed theory of the secular evolution of the mutual orbit due to the
BYORP effect, which predicts that it causes the orbit to expand or contract on a
timescale of thousands of years, as long as the satellite remains synchronous.
Specifically, they applied this theory to the binary near-Earth asteroid 1999 KW4 for
which a detailed model (e.g. mutual orbit and asteroid shapes) is available (Ostro et
al., 2006). From this example, they derived formulae to scale their results for
application to other binary asteroid systems. They showed that the BYORP effect can
be detected in the system by tracking the mean anomaly which changes quadratically in
time for a contracting or expanding mutual orbit.

The mean anomaly of a changing orbit expanded to the second degree in time is expressed by Equations~(\ref{dMd1}) and~(\ref{dMd2}).
$\Delta M_d$ can be expressed using a semimajor axis of the mutual orbit $a$ and its time derivative as
\begin{equation}
\Delta M_d = \frac{1}{2} \dot{n} = - \frac{3 n \dot{a}}{4 a}.
\end{equation}

McMahon and Scheeres (2010b) derived a current semi-major axis expansion rate for 1999 KW$_4$ of $\dot{a}_{\rm KW4} \sim$ 7~cm per year.
Their scaling was adapted by Pravec and Scheirich (2010) to use with parameters that were straightforwardly derived or estimated
from photometric observations for other observed binary asteroids.  The adapted formulas are as follows:
\begin{eqnarray}
\Delta M_d = \frac{K}{\sqrt[3]{q(1+q)} D^2_1 \rho^{\frac{4}{3}} P_{{\rm orb}}^{\frac{2}{3}} a^2_{{\rm hel}} \sqrt{1-e^2_{{\rm hel}}} },\\
K = -24 \pi \left(\frac{3 \pi}{G}\right)^{\frac{4}{3}} \dot{a}_{\rm KW4} \frac{a^3_{\rm KW4} a^2_{\rm hel,KW4} \sqrt{1-e^2_{\rm hel,KW4}}}{(1 + q^{-1}_{\rm KW4}) D^2_{\rm 2,KW4} P^3_{\rm orb,KW4}},\label{EqK}
\end{eqnarray}
where $q$ is the mass ratio between binary components, $D_i$ is the diameter of the $i$-th body, $\rho$ is the bulk density,
$P_{\rm orb}$ is the orbital period, $a_{\rm hel}$ and $e_{\rm hel}$ are the semi-major axis and eccentricity of the binary's heliocentric orbit,
and $G$~is the gravitational constant.

Using the above equations, Pravec and Scheirich (2010) predicted the quadratic drift for several binary near-Earth asteroids and identified seven candidates
for detection of the BYORP effect within years 2010--2015 (including previous observed apparitions of these systems)
using observations with telescopes of sizes about 2 m and smaller: (7088) Ishtar, (65803) Didymos, (66063) 1998 RO$_1$, (88710)	2001 SL$_9$, (137170) 1999 HF$_1$, (175706) 1996 FG$_3$, and (185851) 2000 DP$_{107}$.
The estimated values\footnote{These values of $\Delta M_d$ are based on the shape of 1999 KW$_4$ and they are magnitude, not directional estimates.}
of $\Delta M_d$ for these systems were from $0.24$ to $3.27$ deg/yr$^2$.
The value they predicted for 1996~FG$_3$ was $0.89$ deg/yr$^2$. With the current refined values for the parameters of FG3, the $\Delta M_d$ predicted from
the BYORP theory would be $1.30$~deg/yr$^2$.
Our detected value, $0.04 \pm 0.20$~deg/yr$^2$, is much lower than this estimate.

More recently, Jacobson and Scheeres (2011a) presented an improved theory that incorporates both the BYORP effect and mutual tides between the two components for the first time.
They showed that a stable long-term equilibrium may exist between these two torques if the BYORP effect is removing angular momentum from the orbit.
Since the mean
motion of the orbit is slower than the rotation rate of the primary, the satellite raises a
tidal bulge on the primary that removes energy from the rotation of the primary and
transfers angular momentum to the mutual orbit. These two torques are opposite in
sign and can balance one another because they depend differently on the mutual orbit
semi-major axis. They evolve the mutual orbit to an equilibrium semi-major axis,
where the mutual orbit no longer evolves.


An observation of zero drift in the mean anomaly of the mutual orbit of the binary system
 may indicate the presence of such an equilibrium. Alternatively, the BYORP coefficient
of the secondary could be very small (e.g., for a nearly symmetric secondary)
or the BYORP theory is incorrect. In both these cases, the quadratic drift in the mean anomaly would be dominated
by the tidal expansion of the semi-major axis. Goldreich and Sari (2009) estimate this rate from their rubble pile tidal model
to be 0.012 cm/yr and Taylor and Margot (2011) estimate it to be 0.0044 cm/yr assuming an evolutionary path and timescale.
Respectively, these correspond to $\Delta M_d = -0.0070$~deg/yr$^2$ and $-0.0025$~deg/yr$^2$.
While these drift rates are within the current uncertainties, we consider these alternatives to be unlikely given
the success of detecting evolution from the other radiative torques: Yarkovsky and YORP,
and given the shape model of the secondary of 1999 KW$_4$. Future refinements of the drift of the mean anomaly for 1996 FG$_3$
and future measurements of the drift of the mean anomaly of other singly synchronous binary asteroids will confirm or deny
our interpretation that singly synchronous binary asteroids are in a tidal-BYORP equilibrium.

If the binary
asteroid system is in the equilibrium, then it is possible to learn about the interior
structure of the primary from the semi-major axis and the tidal torque balance. The
BYORP and tidal torques are:

\begin{eqnarray}
\Lambda_B = -\mathcal{B}_2 H_{\rm hel} R_1^3 a,\\
\Lambda_T = \frac{2 \pi k \rho \omega_d^2 R_1^5 q^2}{Q a^6},
\end{eqnarray}

where $\mathcal{B}_2$ is a constant representing an averaged acceleration in the direction parallel to the motion of the secondary (McMahon and Scheeres, 2010b), $H_{\rm hel} = (2/3) F_{\rm hel} / (a_{\rm hel}^2 \sqrt{1-e_{\rm hel}^2})$,
$(2/3) F_{\rm hel} = 6.\bar{6} \times 10^{13}$~kg~km~s$^{-2}$ is the solar constant including a Lambertian factor,
$a_{\rm hel}$ and $e_{\rm hel}$ are heliocentric semimajor axis and eccentricity,
$a$ is measured in primary radii $R_1$,
$k$ and $Q$ are the tidal Love number
and the quality factor of the primary,
$\omega_d = \sqrt{4 \pi \rho G/3}$ is the critical angular frequency, $\rho$ is a density of the body,
and $G$ is the gravitational constant.

The BYORP coefficient $\mathcal{B}_2$ is a function solely of the shape of the secondary and not dependent on the size of the body. Nominally, the
coefficient can have any absolute value between 0 and 2, but it is predicted to be $0.04$ for shapes similar to
1999 KW$_4$ (McMahon and Scheeres, 2010b).\footnote{Jacobson and Scheeres (2011a) defined the dimensionless $\mathcal{B}_2$ to be equal to $\bar{A}_0(2) / R_2^2$,
where $\bar{A}_0(2)$ is a coefficient derived by McMahon and Scheeres (2010a). Since $\bar{A}_0(2)$ is estimated to be $\sim 1.5 \times 10^{-3}$ for 1999 KW$_4$
(McMahon and Scheeres, 2010b), $\mathcal{B}_2$ for 1999 KW$_4$ is $\sim 0.04$. Jacobson and Scheeres (2011a) mistakenly rounded this estimate for $\mathcal{B}_2$ to $10^{-3}.$}

The two tidal parameters $k$ and $Q$ are
not well known for any small body. The tidal Love number $k$ represents how
strongly the spherical harmonic degree two of the gravity potential of the primary responds
to the radial gravitational perturbation of the satellite. More specifically, it is the ratio
of the additional potential produced by the deformation of the primary to the original
gravitational potential. If the primary did not deform at all (i.e. perfectly rigid), the
value of $k$ would be 0, and for a point source perturber, a perfect fluid would have a
value of 3/2 (Murray and Dermott, 1999). The Moon and Earth are estimated to have
tidal Love numbers of 0.03 and 0.3, respectively (Yoder, 1995). The Love number is
expected to decrease with the size of the body as the self-gravity of the body
decreases and the body begins to respond more as a monolithic structure than as a
fluid. How this variation occurs is unknown and complicated by the tidal Love
numbers degeneracy with the tidal parameter $Q$. This parameter is a quality factor
that describes the amount of energy dissipated per cycle over the peak potential
energy stored during the cycle. For small bodies this parameter is often assumed to be
100 (Goldreich and Sari, 2009), although it is expected to depend on the size and
rheology of the body. Taylor and Margot (2011) placed constraints on the tidal
parameters of binary asteroid systems, but these could only be limits to each value
since they had to assume the maximum possible value for the tidal evolution timescale for
each binary system found in synchronous orbit.
They estimated these tidal parameters in terms of the product of the tidal quality factor $Q$
and a rigidity $\mu$, which for solid rock has a value of a few to tens of GPa.
For 1996~FG$_3$, they estimate a $\mu Q = 2.7 \times 10^9$~Pa,
which is approximately $Q/k = 2.7 \times 10^7$.
Outside of these estimates, these tidal parameters are otherwise unknown for small bodies.

In the equilibrium, the torques $\Lambda_B$ and $\Lambda_T$ balance each other and the unknown parameters $\mathcal{B}_2$, $k$ and $Q$
can be directly, though degenerately, determined:
\begin{equation}\label{equality}
\frac{\mathcal{B}_2 Q}{k} = \frac{2 \pi \rho \omega_d^2 R_1^2 q^{4/3}}{H_{\rm hel} a^7}.
\end{equation}

Assuming 1996 FG$_3$ is in the equilibrium state, we obtained a value of $\mathcal{B}_2 Q/k = 2.4 \times 10^3$ for the nominal values of the parameters
on the right-hand side of Eq.~\ref{equality}, with $3\sigma$ uncertainty within a factor of two.
The tidal-BYORP equilibrium provides a method to
directly assess these tidal parameters for the first time given an estimate for the
BYORP coefficient of the secondary $\mathcal{B}_2$.  It is worth emphasizing that $\mathcal{B}_2$ is only a
function of the shape of the secondary, so it is possible to make an estimate of this
value from remote sensing alone. Given radar shape models of the secondary of 1999~
KW$_4$, the best estimate of $\mathcal{B}_2$ is approximately $4 \times 10^{-2}$ (McMahon and Scheeres, 2010b).
Considering that this estimate is based on another asteroid's shape, its
uncertainty is difficult to assess, but McMahon and Scheeres (2013) have estimated
BYORP coefficients from generic asteroid shapes to be between 0 and $5 \times 10^{-2}$.
As a zeroth order estimate, we use $\mathcal{B}_2 = 10^{-2}$ with a factor of five uncertainty.
Therefore, $Q/k = 2.4 \times 10^5$ and with the tidal-BYORP equilibrium it is possible to measure
interior structure properties of the FG3 primary from remote sensing.


The $Q/k$ derived from the tidal-BYORP equilibrium is smaller by a factor of about $10^2$
than the upper limit predicted by Taylor and Margot (2011), who assumed a tidal
evolution timescale of 10 Myr and a mutual orbit that expands its semi-major axis from twice
the primary radius to the current orbit.
Since the tidal evolution timescales are
proportional to $Q/k$, this new value predicts much faster tidal evolution of binary
asteroid systems. Although, the mutual orbit did not necessarily evolve as prescribed
in Taylor and Margot, if it did so with the new tidal parameters, it would only take
approximately $10^5$~years. It's worth noting that the semi-major axis of FG3 is
only 3 primary radii so the evolution assumed in Taylor and Margot has the secondary moving
only 1 primary radii.
In the self-consistent theory put forward by Jacobson and Scheeres (2011a),
the system did not necessarily evolve from a separation of two primary
radii, instead after rotational fission, the mutual orbit stabilized onto
an eccentric orbit with a semi-major axis of a few primary radii.
If that orbit was interior to 3 $R_{\rm 1}$, it would evolve outwards and circularize due to tides and the BYORP effect.
If that orbit was exterior, then it would evolve inwards and circularize due to tides.

Finally, we look at how the result converts to $\mu Q$, the product of the rigidity and quality factor.
Using
\begin{equation}
\mu Q = \frac{4}{19} \frac{Q}{k} G \pi R_1^2 \rho^2
\end{equation}
from Goldreich and Sari (2009), we obtain $\mu Q = 1.3 \times 10^7$~Pa.
While the upper limit on $\mu Q$ of $2.7 \times 10^9$~Pa determined by Taylor and Margot (2011) still maintained the possibility that
1996 FG$_3$ was pre-dominately solid rock if $Q$ was low, now after directly
assessing $\mu Q$, we rule out the possibility that 1996 FG$_3$ is anything but
a `rubble pile' asteroid.  It has a rigidity orders of magnitude below
that of solid rock even given the uncertainties estimating $Q$ and $\mathcal{B}_2$.

The value of $\mu Q$ is derived from the classical theory for monolithic bodies which cannot by applied to a `rubble pile',
but its use allows us to directly compare tidal strengths with older predictions. The value is lower by about four orders of magnitude than estimated for, e.g., tumbling asteroids (see Pravec et al., 2014).
It calls for a re-thinking of the tidal energy dissipation in close asteroid binary systems.  While the present theories of tidal evolution
(see Taylor and Margot, 2011, and references therein) assume an elastic response of the asteroid material to the tidal forces,
the obtained very low $\mu Q$ value may indicate a non-elastic behavior due to the large amplitude of the tidal forcing function.
For instance, as the acceleration vector at the equator
of the FG3 primary changes its direction during rotation of the near-critically spinning primary under the tidal forces from the secondary (see Harris et al., 2009), regolith
particles may move.  Alternatively, a low-load friction theory (a ``tribology theory'') may need to be developed for rubble pile asteroids
as suggested by Jacobson and Scheeres (2011b).

\section{Conclusions}

The near-Earth asteroid (175706)~1996~FG$_3$ is one of the best characterized small asteroid binary systems.  Except for being an apparent
outlier with its primitive composition and related quantities, it is a typical member of the population of near-Earth asteroid binaries for most of its parameters.
With the unique data from our photometric observations taken during its six apparitions over the time interval of almost 17 years, we
constrained the long-term evolution of a small binary asteroid orbit for the first time.
We found the upper limit on drift of the mutual orbit of its components that is consistent with the theory of Jacobson and Scheeres (2011a)
of that synchronous binary asteroids are in a state of stable equilibrium between the BYORP effect and mutual body torques.
The derived material parameters indicate that the primary is a `rubble pile', that a tidal evolution of the system is much faster than estimated before,
and it also calls for a re-thinking of the tidal energy dissipation in close asteroid binary systems.

\bigskip
\textbf{Acknowledgements}

We thank A. W. Harris and the reviewers, J. McMahon and M. \'{C}uk, for their constructive suggestions and comments.
The work at Ond\v{r}ejov was supported by the Grant Agency of the Czech Republic, Grants 205/09/1107 and P209/12/0229, and by program RVO 67985815.
S. Jacobson would like to acknowledge the NASA Earth and Space Science Fellowship as well as thesis support through the
National Optical Astronomy Observatory, which is operated by the Association of
Universities for Research in Astronomy (AURA) under cooperative agreement with
the National Science Foundation. He also would like to acknowledge the assistance of
the staffs of both the Kitt Peak National Observatory and the Apache Point Observatories.
The work of J\v{D} was supported by Charles University in Prague, project PRVOUK P45.
Operations at Carbuncle Hill Observatory and Sugarloaf Mt. Observatory were supported by a Gene Shoemaker NEO grant from the Planetary Society.
D. Polishook is grateful to the AXA research fund for their generous postdoctoral fellowship.
TRAPPIST is a project funded by the Belgian Fund for Scientific Research (Fond National de la
Recherche Scientifique, F.R. SFNRS) under grant FRFC 2.5.594.09.F, with the
participation of the Swiss National Science Foundation (SNF). M. Gillon and
E. Jehin are FNRS Research Associates, J. Manfroid is Research Director FNRS.
The work at Modra was supported by the Slovak Grant Agency for Science VEGA (Grant 1/0670/13).
JL and VAL acknowledges support from the project AYA2012-39115-C03-03 (MINECO).

\bigskip
\textbf{References}

Benner, L. A. M., Brozovic, M., Giorgini, J. D., Lawrence, K. J., Taylor, P. A., Nolan, M. C., Howell, E. S., Busch, M. W., Margot, J. L., Naidu, S. P.,
Magri, C., Shepard, M. K.,
2012. Arecibo and Goldstone Radar Observations of Binary Near-Earth Asteroid and Marco Polo-R Mission Target (175706) 1996 FG3.
ACM 2012, Proce. of the conference held May 16-20, 2012 in Niigata, Japan. LPI Contribution No. 1667, id.6403.

Binzel, R.~P., Harris, A.~W., Bus, S.~J., Burbine, T.~H.,
2001. Spectral properties of near-Earth objects: Palomar and IRTF results for 48 objects including spacecraft targets (9969) Braille and (10302) 1989 ML.
Icarus, 151,  139–-149.

{Binzel}, R.~P., {Perozzi}, E., {Rivkin}, A.~S., {Rossi}, A., {Harris}, A.~W., {Bus}, S.~J., {Valsecchi}, G.~B., {Slivan}, S.~M.,
2004. Dynamical and compositional assessment of near-Earth object mission targets.
Meteoritics and Planetary Science, 39, 351--366.

Binzel, R. P., Polishook, D., DeMeo, F. E., Emery, J. P., Rivkin, A. S.,
2012. Marco Polo-R Target Asteroid (175706) 1996 FG3: Possible Evidence for an Annual Thermal Wave.
43rd Lunar and Planetary Science Conference, held March 19-23, 2012 at The Woodlands, Texas. LPI Contribution No. 1659, id.2222.

Bowell, E., Hapke, B., Domingue, D., Lumme, K., Peltoniemi, J., Harris, A.W.,
1989. Application of photometric models to asteroids.
In: Asteroids II. Univ. Arizona Press, pp. 524–-556.

{Bottke}, Jr., W.~F., {Vokrouhlick{\'y}}, D., {Rubincam}, D.~P., {Nesvorn{\'y}}, D.,
2006. The Yarkovsky and Yorp Effects: Implications for Asteroid Dynamics.
Annual Review of Earth and Planetary Sciences, 34, 157--191.

{Brosch}, N., {Polishook}, D., {Shporer}, A., {Kaspi}, S., {Berwald}, A., {Manulis}, I.,
2008. The Centurion 18 telescope of the Wise Observatory.
Astrophysics and Space Science, 314, 163--176.

{Chesley}, S.~R., {Ostro}, S.~J., {Vokrouhlick{\'y}}, D., {{\v C}apek}, D., {Giorgini}, J.~D., {Nolan}, M.~C., {Margot}, J.-L., {Hine}, A.~A., {Benner}, L.~A.~M., {Chamberlin}, A.~B.,
2003. Direct Detection of the Yarkovsky Effect by Radar Ranging to Asteroid 6489 Golevka.
Science, 302, 1739--1742.

\'{C}uk, M., Burns, J.A.,
2005. Effects of thermal radiation on the dynamics of binary NEAs.
Icarus, 176, 418--431.

{de Le{\'o}n}, J., {Moth{\'e}-Diniz}, T., {Licandro}, J., {Pinilla-Alonso}, N., {Campins}, H.,
2011. New observations of asteroid (175706) 1996 FG3, primary target of the ESA Marco Polo-R mission.
Astron. and Astroph., 530, L12.

{de Le{\'o}n}, J., {Lorenzi}, V., {Al{\'{\i}}-Lagoa}, V., {Licandro}, J., {Pinilla-Alonso}, N., {Campins}, H.,
2013. Additional spectra of asteroid 1996 FG3, backup target of the ESA MarcoPolo-R mission.
Astron. and Astroph., 556, A33.

{Durda}, D.~D., {Bottke}, W.~F., {Enke}, B.~L., {Merline}, W.~J., {Asphaug}, E., {Richardson}, D.~C., {Leinhardt}, Z.~M.
2004. The formation of asteroid satellites in large impacts: results from numerical simulations.
Icarus, 167, 382--396.

Gal\'ad, A., Pravec, P., Gajdo\v{s}, \v{S}., Korno\v{s}, L., Vil\'agi, J.,
2007. Seven asteroids studied from Modra observatory in the course of binary asteroid photometric campaign.
Earth Moon Planets 101, 17–-25.

Goldreich, P., Sari, R.,
2009. Tidal Evolution of Rubble Piles.
Astroph. Journal, 691, 54--60.

Harris, A.W., Lupishko, D.F.,
1989. Photometric lightcurve observations and reduction techniques. In Asteroids II (R. Binzel, T. Gehrels, and M.
Matthews, eds.), Tucson: Univ. of Arizona Press, pp 39--53.

Harris, A.W., Fahnestock, E.G., Pravec, P.,
2009. On the shapes and spins of ``rubble pile'' asteroids.
Icarus 199, 310--318.

Jacobson, S.A., Scheeres, D.J.,
2011a. Long-term Stable Equilibria for Synchronous Binary Asteroids.
ApJ Letters, 736, L19.

Jacobson, S.A., Scheeres, D.J., 2011b.  A long-term stable equilibrium for synchronous binaries including tides and the BYORP effect.
AAS/Division on Dynamical Astronomy Meeting Abstracts 42, Talk: 01.02.

Kaasalainen, M., Torppa, J.,
2001. Optimization Methods for Asteroid Lightcurve Inversion. I. Shape Determination.
Icarus, 153, 24--36.

Kaasalainen, M., Torppa, J., and Muinonen, K.,
2001. Optimization Methods for Asteroid Lightcurve Inversion II. The Complete Inverse Problem.
Icarus, 153, 37--51.

Kaasalainen, M., Mottola, S., Fulchignoni, M.,
2002. Asteroid Models from Disk-integrated Data.
In Asteroids III, ed. W. F. Bottke Jr., A. Cellino, P. Paolicchi, R. P. Binzel, University of Arizona Press, Tucson, pp. 139.

{Kaasalainen}, M., {{\v D}urech}, J., {Warner}, B.~D., {Krugly}, Y.~N., {Gaftonyuk}, N.~M.,
2007. Acceleration of the rotation of asteroid 1862 Apollo by radiation torques.
Nature, 446, 420-422.

Krugly, Y. N., Belskaya, I. N., Shevchenko, V. G., et al.,
2002. The Near-Earth Objects Follow-up Program. IV. CCD Photometry in 1996-1999.
Icarus, 158, 294--304.

{Lowry}, S.~C., {Fitzsimmons}, A., {Pravec}, P., {Vokrouhlick{\'y}}, D., {Boehnhardt}, H., {Taylor}, P.~A., {Margot}, J.-L., {Gal{\'a}d}, A., {Irwin}, M., {Irwin}, J., {Kusnir{\'a}k}, P.,
2007. Direct Detection of the Asteroidal YORP Effect.
Science, 316, 272--274.

McMahon, J., Scheeres, D.,
2010a. Secular orbit variation due to solar radiation effects: a detailed model for BYORP.
Celestial Mechanics and Dynamical Astronomy, 106, 261--300.

McMahon, J., Scheeres, D.,
2010b. Detailed prediction for the BYORP effect on binary near-Earth Asteroid (66391) 1999 KW4 and implications for the binary population.
Icarus, 209, 494--509.

{McMahon}, J.~W., {Scheeres}, D.,
2013. A Statistical Analysis of YORP Coefficients.
45th annual meeting of the Division for Planetary Sciences of the American Astronomical Society, Oct. 6 - Oct. 11,  2013, Denver, CO.

Mottola, S., Lahulla, F.,
2000. Mutual Eclipse Events in Asteroidal Binary System 1996~FG$_3$: Observations and a Numerical Model.
Icarus, 146, 556--567.

Mueller, M., Delbo', M., Hora, J. L., Trilling, D. E., Bhattacharya, B., Bottke, W. F., Chesley, S., Emery, J. P., Fazio, G., Harris, A. W., Mainzer, A., Mommert, M., Penprase, B., Smith, H. A., Spahr, T. B., Stansberry, J. A., Thomas, C. A.,
2011. ExploreNEOs. III. Physical Characterization of 65 Potential Spacecraft Target Asteroids.
Astronom. Journal, 141, 109.

Murray, C.D., Dermott, S.F.,
1999. Solar System Dynamics. Cambridge University Press.


Ostro, S.J., Margot, J.-L., Benner, L.A.M., Giorgini, J.D., Scheeres, D.J., Fahnestock, E.G., Broschart, S.B., Bellerose, J., Nolan, M.C., Magri, C., Pravec, P., Scheirich, P., Rose, R., Jurgens, R.F., De Jong, E.M., Suzuki, S.,
2006. Radar Imaging of Binary Near-Earth Asteroid (66391) 1999 KW4.
Science, 314, 1276--1280.

Perna, D., Dotto, E., Barucci, M. A., Fornasier, S., Alvarez-Candal, A., Gourgeot, F., Brucato, J. R., Rossi, A.,
2013. Ultraviolet to near-infrared spectroscopy of the potentially hazardous, low delta-V asteroid (175706) 1996 FG3. Backup target of the sample return mission MarcoPolo-R.
Astron. and Astroph., 555, A62.

{Pilcher}, F., {Briggs}, J.~W., {Franco}, L., {Inasaridze}, R.~Y., {Krugly}, Y.~N., {Molotiv}, I.~E., {Klinglesmith}, III, D.~A., {Pollock}, J., {Pravec}, P.,
2012. Rotation Period Determination for 5143 Heracles.
Minor Planet Bulletin, 39, 148--151.

Polishook, D., Brosch, N.,
2009. Photometry and spin rate distribution of small-sized main belt asteroids.
Icarus, 199, 319--332.

Popescu, M., Birlan, M., Nedelcu, D. A.,
2012. Modeling of asteroid spectra - M4AST.
Astron. and Astroph., 544, A130.

Pravec, P., \v{S}arounov\'a, L, Wolf, M., Mottola, S., Lahulla, F.,
1998. 1996 FG3. IAUC 7074.

Pravec, P., \v{S}arounov\'a, L., Rabinowitz, D.L., Hicks, M.D., Wolf, M., Krugly, Y.N., Velichko, F.P., Shevchenko, V.G., Chiorny, V.G., Gaftonyuk, N.M., Genevier, G.,
2000. Two-Period Lightcurves of 1996~FG$_3$, 1998 PG, and (5407) 1992 AX: One Probable and Two Possible Binary Asteroids.
Icarus, 146, 190--203.

Pravec, P., Scheirich, P., Ku\v{s}nir\'ak, P., \v{S}arounov\'a, L., Mottola, S., Hahn, G., Brown, P., Esquerdo, G., Kaiser, N., Krzeminski, Z., Pray, D.P., Warner, B.D., Harris, A.W., Nolan, M.C., Howell, E.S., Benner, L.A. M., Margot, J.-L., Gal\'ad, A., Holliday, W., Hicks, M.D., Krugly, Yu.N., Tholen, D., Whiteley, R., Marchis, F., Degraff, D.R., Grauer, A., Larson, S., Velichko, F.P., Cooney, W.R., Stephens, R., Zhu, J., Kirsch, K., Dyvig, R., Snyder, L., Reddy, V., Moore, S., Gajdo\v{s}, \v{S}., Vil\'agi, J., Masi, G., Higgins, D., Funkhouser, G., Knight, B., Slivan, S., Behrend, R., Grenon, M., Burki, G., Roy, R., Demeautis, C., Matter, D., Waelchli, N., Revaz, Y., Klotz, A., Rieugn\'e, M., Thierry, P., Cotrez, V., Brunetto, L., Kober, G.,
2006. Photometric survey of binary near-Earth asteroids.
Icarus, 181, 63--93.

Pravec, P., Scheirich, P.,
2010. Binary System Candidates for Detection of BYORP.
42nd annual meeting of the Division for Planetary Sciences of the American Astronomical Society, Oct. 3 - Oct. 8,  2010, Pasadena, CA.

Pravec, P., Harris, A.W.,
2007. Binary asteroid population. 1: Angular momentum content.
Icarus 190, 250–-259.

{Pravec}, P., {Vokrouhlick{\'y}}, D., {Polishook}, D., {Scheeres}, D.~J., {Harris}, A.~W., {Gal{\'a}d}, A., {Vaduvescu}, O., {Pozo}, F., {Barr}, A., {Longa}, P., {Vachier}, F., {Colas}, F., {Pray}, D.~P., {Pollock}, J., {Reichart}, D., {Ivarsen}, K., {Haislip}, J., {Lacluyze}, A., {Ku{\v s}nir{\'a}k}, P., {Henych}, T., {Marchis}, F.,	{Macomber}, B., {Jacobson}, S.~A., {Krugly}, Y.~N., {Sergeev}, A.~V., {Leroy}, A.,
2010. Formation of asteroid pairs by rotational fission.
Nature, 466, 1085--1088.

Pravec, P., Scheirich, P., Vokrouhlick\'y, D., Harris, A. W., Ku\v{s}nir\'ak, P., Hornoch, K., Pray, D. P., Higgins, D., Gal\'ad, A., Vil\'agi, J., Gajdo\v{s}, \v{S}., Korno\v{s}, L., Oey, J., Hus\'{a}rik, M., Cooney, W. R., Gross, J., Terrell, D., Durkee, R., Pollock, J., Reichart, D. E., Ivarsen, K., Haislip, J., LaCluyze, A., Krugly, Yu. N., Gaftonyuk, N., Stephens, R. D., Dyvig, R., Reddy, V., Chiorny, V., Vaduvescu, O., Longa-Pena, P., Tudorica, A., Warner, B. D., Masi, G., Brinsfield, J., Gon\c{c}alves, R., Brown, P., Krzeminski, Z., Gerashchenko, O., Shevchenko, V., Molotov, I., Marchis, F.,
2012. Binary asteroid population. 2. Anisotropic distribution of orbit poles of small, inner main-belt binaries.
Icarus, 218, 125--143.

{Pravec}, P., {Scheirich}, P., {{\v D}urech}, J., {Pollock}, J., {Ku{\v s}nir{\'a}k}, P., {Hornoch}, K., {Gal{\'a}d}, A., {Vokrouhlick{\'y}}, D., {Harris}, A.~W., {Jehin}, E.,	{Manfroid}, J., {Opitom}, C., {Gillon}, M., {Colas}, F., {Oey}, J., {Vra{\v s}til}, J., {Reichart}, D., {Ivarsen}, K., {Haislip}, J., {LaCluyze}, A.
2014. The tumbling spin state of (99942) Apophis.
Icarus, 233, 48--60.

Rayner, J. T., Toomey, D. W., Onaka, P. M., Denault, A. J., Stahlberger, W. E., Vacca, W. D., Cushing, M. C., and Wang, S.,
2003. SpeX: A Medium-Resolution 0.8-5.5 micron Spectrograph and Imager for the NASA Infrared Telescope Facility.
PASP 115, 362.

{Rivkin}, A.~S., {Howell}, E.~S., {Vervack}, R.~J., {Magri}, C., {Nolan}, M.~C., {Fernandez}, Y.~R., {Cheng}, A.~F., {Antonietta Barucci}, M., {Michel}, P.,
2013. The NEO (175706) 1996 FG3 in the 2-4 µm spectral region: Evidence for an aqueously altered surface.
Icarus, 223, 493--498.

Scheirich, P., Pravec, P.,
2009. Modeling of lightcurves of binary asteroids.
Icarus, 200, 531--547.

{Taylor}, P.~A., {Margot}, J.-L., {Vokrouhlick{\'y}}, D., {Scheeres}, D.~J., {Pravec}, P., {Lowry}, S.~C., {Fitzsimmons}, A., {Nolan}, M.~C., {Ostro}, S.~J., {Benner}, L.~A.~M., {Giorgini}, J.~D., {Magri}, C.,
2007. Spin Rate of Asteroid (54509) 2000 PH5 Increasing Due to the YORP Effect.
Science, 316, 274--277.

Taylor, P.A., Margot, J.-L.,
2011. Binary asteroid systems: Tidal end states and estimates of material properties.
Icarus, 212, 661--676.

{Trigo-Rodr{\'{\i}}guez}, J.~M., {Garc{\'{\i}}a-Hern{\'a}ndez}, D.~A., {S{\'a}nchez}, A., {Lacruz}, J., {Davidsson}, B.~J.~R., {Rodr{\'{\i}}guez}, D., {Pastor}, S., {de Los Reyes}, J.~A.,
2010. Outburst activity in comets - II. A multiband photometric monitoring of comet 29P/Schwassmann-Wachmann 1.
Month. Not. of the Royal Astr. Soc., 409, 1682--1690.

{Walsh}, K.~J., {Delbo'}, M., {Mueller}, M., {Binzel}, R.~P., {DeMeo}, F.~E.,
2012. Physical Characterization and Origin of Binary Near-Earth Asteroid (175706) 1996 FG$_{3}$.
Astroph. Journal, 748, 104.

Wolters, S.~D., Green, S.~F.,
2009. Investigation of systematic bias in radiometric diameter determination of near-Earth asteroids: the night emission simulated thermal model (NESTM).
Month. Not. of the Royal Astr. Soc., 400, 204--218.

{Wolters}, S.~D., {Rozitis}, B., {Duddy}, S.~R., {Lowry}, S.~C., {Green}, S.~F., {Snodgrass}, C., {Hainaut}, O.~R., {Weissman}, P.,
2011. Physical characterization of low delta-V asteroid (175706) 1996 FG3.
Month. Not. of the Royal Astr. Soc., 418, 1246--1257.

Warner, B.D., Pray, D.P.,
2009. Analysis of the Lightcurve of (6179) Brett.
Minor Planet Bull. 36, 166-–168.

Warner, B.D., Stephens, R. D., Pray, D. P.,
2013. Lightcurve Analysis of the Near Earth Asteroid (138095) 2000 DK79.
Minor Planet Bull. 41, 75--77.

Yoder, C.~F.,
1995. Astrometric and Geodetic Properties of Earth and the Solar System.
Global earth physics a handbook of physical constants. Ed. by T. J. Ahrens. ISBN 0-87590-851-9, American Geophysical Union, Washington, DC.

Yu, L., Ji, J., Wang, S.,
2014. Shape, thermal and surface properties determination of a candidate spacecraft target asteroid (175706) 1996 FG3.
Monthly Notices of the Royal Astr. Soc., 439, 3357--3370.

\end{document}